\documentclass[sigconf]{acmart}

\usepackage{booktabs} 

\usepackage[linesnumbered,boxed,ruled,commentsnumbered]{algorithm2e}
\usepackage{times}  
\usepackage{helvet}  
\usepackage{courier}  
\usepackage{url}  
\usepackage{graphicx}  

\usepackage{multirow}
\usepackage{subfigure}
\usepackage{amsmath,bm}
\usepackage{enumerate}

\usepackage{times}
\usepackage{helvet}
\usepackage{courier}
\setcopyright{rightsretained}

\acmDOI{10.475/123_4}

\acmISBN{123-4567-24-567/08/06}

\acmConference[WWW 2019]{WWW 2019}{May 2019}{San Francisco, USA}
\acmYear{2019}
\copyrightyear{2019}

\acmArticle{4}
\acmPrice{15.00}

\editor{Jennifer B. Sartor}
\editor{Theo D'Hondt}
\editor{Wolfgang De Meuter}

\begin{document}
\title{Heterogeneous Graph Attention Network}

\author{Xiao Wang, Houye Ji}
\affiliation{%
  \institution{Beijing University of Posts and Telecommunications}
  \streetaddress{P.O. Box 1212}
  \city{Beijing}
  \state{China}
  \postcode{43017-6221}
}
\email{  {xiaowang,jhy1993}@bupt.edu.cn}

%
\author{Chuan Shi*, Bai Wang}
\affiliation{%
	\institution{Beijing University of Posts and Telecommunications}
	\streetaddress{P.O. Box 1212}
	\city{Beijing}
	\state{China}
	\postcode{43017-6221}
}
\email{{shichuan,wangbai}@bupt.edu.cn}

\author{Peng Cui, P. Yu}
\affiliation{%
	\institution{Tsinghua University}
	\streetaddress{8600 Datapoint Drive}
	\city{Beijing}
	\state{China}
	\postcode{78229}}
\email{{cuip, psyu}@tsinghua.edu.cn}

\author{Yanfang Ye}
\affiliation{%
 \institution{West Virginia University}
 \streetaddress{Rono-Hills}
 \state{WV}
 \country{USA}}
\email{yanfang.ye@mail.wvu.edu}

\renewcommand{\shortauthors}{Xiao Wang, Houye Ji, Chuan Shi, Bai Wang et al.}

\begin{abstract}
	
	Graph neural network, as a powerful graph representation technique based on deep learning, has shown superior performance and attracted considerable research interest. However, it has not been fully considered in graph neural network for heterogeneous graph which contains different types of nodes and links. The heterogeneity and rich semantic information bring great challenges for designing a graph neural network for heterogeneous graph.
	Recently, one of the most exciting advancements in deep learning is the attention mechanism, whose great potential has been well demonstrated in various areas.
	In this paper, we first propose a  novel heterogeneous graph neural network based on the hierarchical attention, including node-level and semantic-level attentions. Specifically,
	the node-level attention aims to learn the importance between a node and its meta-path based neighbors, while the semantic-level attention is able to learn the importance of different meta-paths. With the learned importance from both node-level and semantic-level attention, the importance of node and meta-path 
	can be fully considered. Then  the proposed model can generate node embedding by aggregating features  from meta-path based neighbors in a hierarchical manner. Extensive experimental results on three real-world heterogeneous graphs not only show the superior performance of our proposed model over the state-of-the-arts, but also demonstrate its potentially good interpretability for graph analysis.
\end{abstract}

%

%
%


\maketitle

\section{Introduction}
The real-world data usually come together with the graph structure, 
such as social networks, citation networks, and the world wide web. Graph neural network (GNN), 
as a powerful deep representation learning method for such graph data, 
has shown superior performance on network analysis and aroused considerable research interest. 
For example, \cite{gnn05,gnn09,ggnn}
leverage deep neural network to learn node representations based on node features 
and the graph structure. 
Some works \cite{16chebyshev,gcn,graphsage} 
propose the graph convolutional networks by generalizing the convolutional operation to graph.
A recent research trend in deep learning is the attention mechanism,
which deals with variable sized data and encourages the model to focus on the most salient parts of data. 
It has demonstrated the effectiveness in deep neural network framework and is widely applied to various applications, such as text analysis \cite{bahdanau2014neural}, knowledge graph \cite{schlichtkrull2018modeling} and image processing \cite{show_xu2015show}.
Graph Attention Network (GAT) \cite{gat}, a novel convolution-style graph neural network, leverages attention mechanism for the homogeneous graph which includes only one type of nodes or links.

\begin{figure}
	\centering
	\includegraphics[width=1\columnwidth, height=4cm]{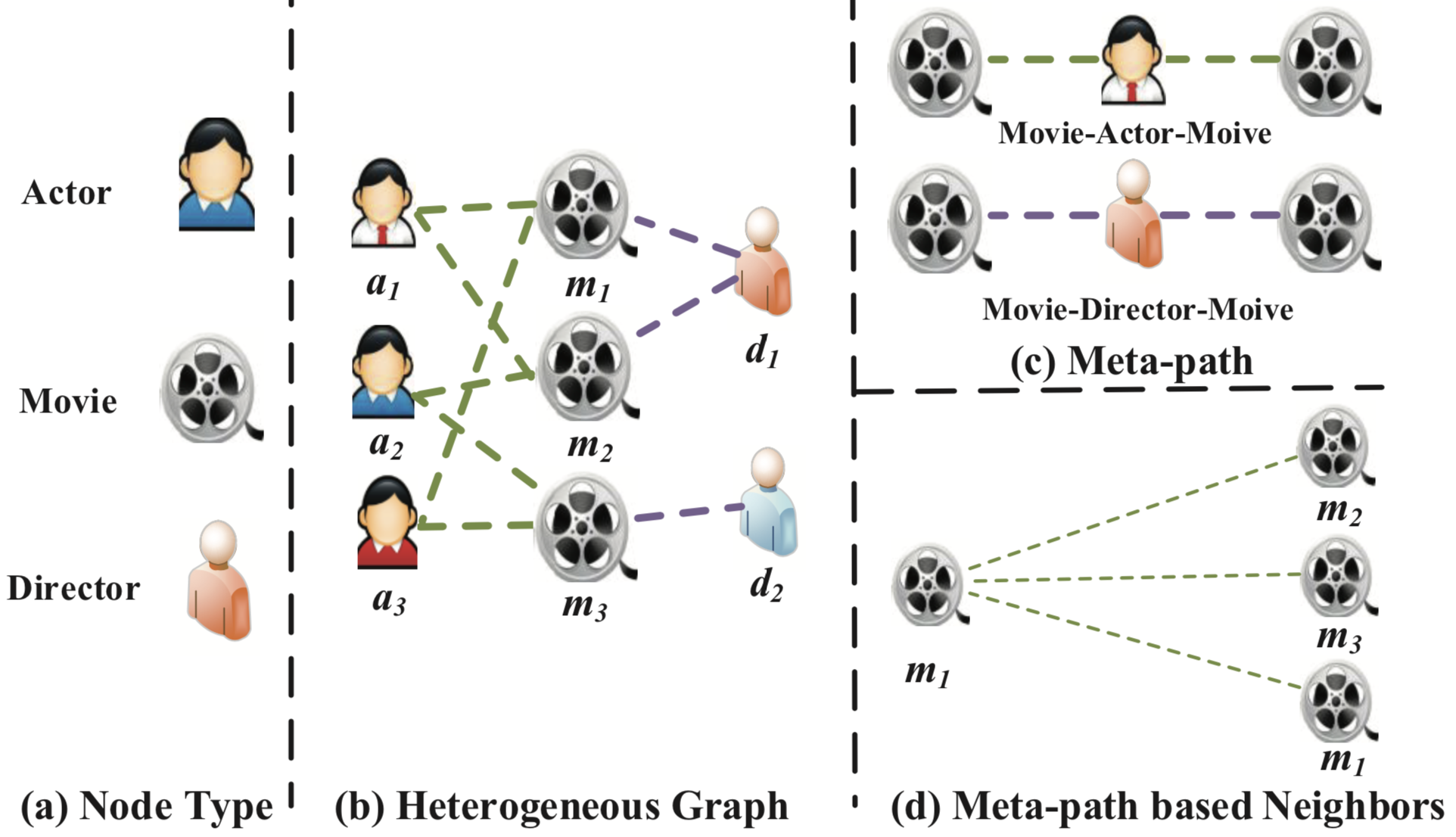}
	\caption{
		An illustrative example of a heterogenous graph (IMDB). 
		(a) Three types of nodes (i.e., actor, movie, director).
		(b) A heterogenous graph IMDB  consists three types of nodes  and two types of connections.
		(c) Two meta-paths involved in IMDB (i.e., Moive-Actor-Moive and Movie-Director-Movie). (d) Moive $m_1$ and its meta-path based neighbors (i.e., $m_1$, $m_2$ and $m_3$). 
	}
	\label{fig_hin}
\end{figure}

Despite the success of attention mechanism in deep learning, it has not been considered in the graph neural network framework for heterogeneous graph. As a matter of fact, the real-world graph usually comes with multi-types of nodes and edges, 
also widely known as heterogeneous information network (HIN) \cite{Shi2017ASO}. For convenience, we uniformly call it heterogeneous graph in this paper.
Because the heterogeneous graph contains more comprehensive information and rich semantics, it has been widely used in many data mining tasks. Meta-path \cite{sun2011pathsim}, a composite relation connecting two objects, is a widely used structure to capture the semantics. 
Taking the movie data IMDB\footnote{https://www.imdb.com} shown in Figure \ref{fig_hin}(a) as an example, it contains three types of nodes include movie, actor and director.
A relation between two movies can be revealed by meta-path Movie-Actor-Movie (\emph{MAM}) which describes the co-actor relation,
while Movie-Director-Movie (\emph{MDM}) means that they are directed by the same director.
As can be seen, depending on the meta-paths,
the relation between nodes in the heterogeneous graph can have different semantics.
Due to the complexity of heterogeneous graph, traditional graph neural network cannot be directly applied to heterogeneous graph.

Based on the above analysis, when designing graph neural network architecture with attention mechanism for heterogeneous graph, 
we need to address the following new requirements.

\textbf{Heterogeneity of graph.}
The heterogeneity is an intrinsic property of heterogeneous graph, i.e., various types of nodes and edges.
For example, different types of nodes have different traits and their features may fall in different feature space. Still taking IMDB as an example, the feature of an actor may involve in sex, age and nationality. On the other hand, the feature of movie may involve in plot and actors. 
How to handle such complex structural information and  preserve the diverse feature information simultaneously is an urgent problem that needs to be solved. 

\textbf{Semantic-level attention.} Different meaningful and complex semantic information are involved in heterogeneous graph, which are usually reflected by meta-paths \cite{sun2011pathsim}. 
Different meta-paths in heterogeneous graph may extract diverse semantic information. How to select the most meaningful meta-paths and fuse the semantic information for the specific task is an open problem \cite{schain,chen2017task,Shang2016MetaPathGE}.
Semantic-level attention aims to learn the importance of each meta-path and assign proper weights to them. Still taking IMDB as an example, \emph{The Terminator} can either connect to  \emph{The Terminator 2} via  Movie-Actor-Movie (both starred by \emph{Schwarzenegger}) or connect to \emph{Birdy} via Movie-Year-Movie (both shot in \emph{1984}). 
However, when identifying the genre of the movie \emph{The Terminator}, \emph{MAM} usually plays more important role, rather than \emph{MYM}.
Therefore, treating different meta-paths equally is unpractical and will weaken the semantic information provided by some useful meta-paths.

\textbf{Node-level attention.}
In a heterogeneous graph, nodes can be connected via various types of relation, e.g., meta-path.
Given a meta-path, each node has lots of meta-path based neighbors. How to distinguish the subtle difference of there neighbors and select some informative neighors is required.
For each node, node-level attention aims to learn the importance of meta-path based neighbors and assign different attention values to them.
Still taking IMDB as an example, 
when using the meta-path Movie-Director-Moive (the movies are with the same director), \emph{The Terminator}  will connect to \emph{Titanic} and \emph{The Terminator 2}  via director \emph{James Cameron}.
To better identify the genre of \emph{The Terminator} as sci-fi movie, the model should pay more attention to \emph{The Terminator 2}, rather than \emph{Titanic}. Therefore, how to design a model which can discover the subtle differences of neighbors and learn their weights properly will be desired.



In this paper, 
we propose a novel 
\textbf{H}eterogeneous graph \textbf{A}ttention \textbf{N}etwork, named HAN, which considers both of node-level and semantic-level attentions.
In particular, given the node features as input, we use the type-specific transformation matrix to project different types of node features into the same space.
Then the node-level attention is able to learn the attention values between the nodes and their meta-path based neighbors, while the semantic-level attention aims to learn the attention values of different meta-paths for the specific task in the heterogeneous graph. Based on the learned attention values in terms of the two levels, our model can get the optimal combination of neighbors and multiple meta-paths in a hierarchical manner, which enables the learned node embeddings to better capture the complex structure and rich semantic information  in a heterogeneous graph.
After that, 
the overall model can be optimized via backpropagation in an end-to-end manner.

The contributions of our work are summarized as follows:

\textbullet\  To our best knowledge, this is the first attempt to study the heterogeneous graph neural network based on attention mechanism. Our work enables the graph neural network to be directly applied to the heterogeneous graph, and further facilitates the heterogeneous graph based applications.
	
\textbullet\   We propose a novel heterogeneous graph attention network (HAN) which includes both of the node-level and semantic-level attentions.
	Benefitting from such hierarchical attentions, the proposed HAN can take the importance of nodes and meta-paths into consideration simultaneously. Moreover, our model is high efficiency, with the linear complexity with respect to the number of meta-path based node pairs, which can be applied to large-scale heterogeneous graph.
	
\textbullet\  	We conduct extensive experiments to evaluate the performance of the proposed model. 
	The results show the superiority of the proposed model by comparing with the state-of-the-art models.
	More importantly, by analysing 
	the hierarchical attention mechanism, the proposed HAN
	demonstrates its potentially good interpretability for heterogeneous graph analysis.

\section{Related Work}
\subsection{Graph Neural Network}
Graph neural networks (GNNs) which aim to extend the deep neural network to deal with arbitrary graph-structured data are introduced in \cite{gnn05,gnn09}. Yujia Li et al. \cite{ggnn} proposes a propagation model which can incorporate gated recurrent units to propagate information across all nodes. Recently, there is a surge of generalizing convolutional operation on the graph-structured data. The graph convolutional neural work generally falls into two categories, namely spectral domain and non-spectral domain. On one hand, spectral approaches work with a spectral representation of the graphs. Joan Bruna et al. \cite{gft14iclr} extends convolution to general graphs by finding the corresponding Fourier basis. 
Micha{\"e}l et al. \cite{16chebyshev} utilizes
K-order Chebyshev polynomials to approximate smooth filters in the spectral domain. Kipf et al. \cite{gcn} proposes a spectral approach, named Graph Convolutional Network,
which designs a graph
convolutional network via a localized first-order approximation of spectral graph convolutions. 
On the other hand, we also have non-spectral approaches, which define convolutions directly on the graph, operating on groups of spatially close neighbors. Hamilton et al. \cite{graphsage} introduces GraphSAGE which 
performs a neural network based aggregator over a fixed size node neighbor. It can learn a function that generates embeddings by aggregating features from a node’s local neighborhood.


%

Attention mechanisms, e.g.,
self-attention  \cite{attisallyouneed} and soft-attention \cite{bahdanau2014neural}, 
have become one of the most influential mechanisms in deep learning. Some previous works introduce the attention mechanism for graph based applications, e.g., the recommendation \cite{mcrec,han2018aspect}. 
Inspired by attention mechanism, 
Graph Attention Network \cite{gat} is proposed to learn the importance between nodes and its neighbors and fuse the neighbors to 
perform node classification. 
However, the above graph neural network cannot deal with various types of nodes and edges and can only be applied to the homogeneous graphs.

\subsection{Network Embedding}

Network embedding, i.e., network representation learning (NRL), is proposed to embed network into a low dimensional space while preserving the network structure and property so that the learned embeddings can be applied to the downstream network tasks.
For example, the random walk based methods \cite{perozzi2014deepwalk,grover2016node2vec}, 
the deep neural network based methods \cite{wang2016structural}, the matrix factorization based methods \cite{hope,mnmf}, and others, e.g., LINE \cite{Tang2015LINELI}. However, all these algorithms are proposed for the homogeneous graphs. Some elaborate reviews
can be found in \cite{cui2018survey,goyal2017graph}.

Heterogeneous graph embedding mainly focuses on preserving the meta-path based structural information.
ESim \cite{Shang2016MetaPathGE} accepts user-defined meta-paths as guidance to learn vertex vectors in a user-preferred embedding space for similarity search. 
Even through ESim can utilize multiple meta-paths, it cannot learn the importance of meta-paths. To achieve the best performance, ESim needs to conduct grid search to find the optimal weights of hmeta-paths. It is pretty hard to find the optimal combination for specific task.
Metapath2vec \cite{Dong2017metapath2vecSR} designs a meta-path based random walk 
and utilizes skip-gram to perform heterogeneous graph embedding. However, metapath2vec can only utilize one meta-path and may ignore some useful information.
Similar to metapath2vec, HERec \cite{HERec} proposes a type constraint strategy to filter the node sequence and capture the complex semantics reflected in heterogeneous graph.
HIN2Vec \cite{Fu2017HIN2VecEM} carries out multiple prediction training tasks which learn the latent vectors of nodes and meta-paths simultaneously. Chen et al. \cite{chen2018pme} proposes a projected metric embedding model, named PME, which can preserve node proximities via Euclidian Distance. 
PME projects different types of node into the same relation space and conducts
heterogeneous link prediction.
To study the problem of comprehensive describe heterogeneous graph, Chen et al. \cite{shi2018easing} proposes HEER which can embed heterogeneous graph via edge representations. Fan et al. \cite{fan2018gotcha} proposes a embedding model metagraph2vec, where both the structures and semantics are maximally preserved for malware detection.
Sun et al. \cite{sun2018joint} proposes meta-graph-based network embedding models, which simultaneously considers
the hidden relations of all meta information of a meta-graph.
In summary, all these aforementioned algorithms do not consider the attention mechanism in heterogeneous graph representation learning.






\section{Preliminary}
A heterogeneous  graph is a special kind of information network, which contains either multiple types of objects or multiple types of links.

\begin{definition}{\textbf{ Heterogeneous  Graph }\cite{sun2013mining}.}
	A  heterogeneous graph, denoted as $\mathcal{G}=(\mathcal{V},\mathcal{E})$, consists of an object set $\mathcal{V}$ and a link set $\mathcal{E}$.
	A heterogeneous graph is also associated with a node type mapping function $\phi:\mathcal{V}\rightarrow \mathcal{A}$ and 
	a link type mapping function $\psi: \mathcal{E}\rightarrow \mathcal{R}$. $\mathcal{A}$ and $\mathcal{R}$ denote 
	the sets of predefined object types and link types, where $|\mathcal{A}|+|\mathcal{R}|>2$.
\end{definition}

\textbf{Example}. As shown in Figure \ref{fig_hin}(a), we construct a heterogeneous graph to model the IMDB. It consists of multiple types of objects ( Actor (A), Movie (M), Director (D))  and relations (shoot relation between movies and directors, role-play relation between actors and movies). 
%


In heterogeneous graph, two objects can be connected via different semantic paths, which are called meta-paths.
%
\begin{definition}{\textbf{Meta-path }\cite{sun2011pathsim}.}
	A meta-path $\Phi$ is defined as a path in the form of $A_1 \xrightarrow{R_1} A_2 \xrightarrow{R_2} \cdots \xrightarrow{R_l} A_{l+1}$ (abbreviated as $A_1A_2 \cdots A_{l+1}$), which describes a composite relation $R = R_1 \circ R_2 \circ \cdots \circ R_l$ between objects $A_1$ and $A_{l+1}$, where $\circ$ denotes the composition operator on relations.
\end{definition}

\textbf{Example}. As shown in Figure \ref{fig_hin}(a), two movies can be connected via multiple meta-paths, e.g., Movie-Actor-Movie (\emph{MAM}) and Movie-Director-Movie (\emph{MDM}). Different meta-paths always reveal different semantics. For example, the \emph{MAM} means the co-actor relation, while Movie-Director-Movie (\emph{MDM}) means they are directed by the same director.

Given a meta-path $\Phi$, there exists a set of meta-path based neighbors of each node which can reveal diverse structure information and rich semantics in a heterogeneous graph. 
\begin{definition}\textbf{Meta-path based Neighbors.}
	Givien a node $i$ and a meta-path $\Phi$ in a heterogeneous graph, the meta-path based neighbors $\mathcal{N}_i^{\Phi}$ of node $i$ are defined as the set of nodes which connect with node $i$ via meta-path $\Phi$. Note that the node's neighbors includes itself.
\end{definition}
\textbf{Example}. Taking Figure \ref{fig_hin}(d) as an example, given the meta-path Movie-Actor-Movie, the meta-path based neighbors of $m_1$ includes $m1$ (itself), $ m_2$ and $m_3$. Similarly, the neighbors of $m_1$ based on meta-path Movie-Director-Movie includes $m_1$ and  $m_2$. Obviously, meta-path based neighbors can exploit different aspects of structure information in heterogeneous graph. We can get meta-path based neighbors by
the multiplication of a sequences of adjacency matrices.

Graph neural network has been proposed to deal with arbitrary graph-structured data. However, all of them are designed for homogeneous network \cite{gcn,gat}. Since meta-path and meta-path based neighbors are two fundamental structures in a heterogeneous graph, next, we will present a novel graph neural network for heterogeneous graph data, which is able to exploit the subtle difference of nodes and meta-paths. The notations we will use throughout the article are summarized in Table \ref{tab_notation}.

\begin{table}
	\caption{Notations and Explanations.}
	\label{tab_notation}
	\begin{tabular}{ccl}
		\toprule
		Notation&Explanation\\
		\midrule
		
		
		${\Phi}$ & Meta-path \\
		${\mathbf{h}}$& Initial node feature \\
		$\mathbf{M}_{\phi}$ & Type-specific transformation matrix \\
		$\mathbf{h}' $& Projected node feature \\
		$e_{ij}^{\Phi}$ & Importance of meta-path based node pair ($i$,$j$) \\
		$\mathbf{a}_{\Phi}$ & Node-level attention vector for meta-path $\Phi$ \\
		$\alpha_{ij}^{\Phi}$ & Weight of 
		meta-path based node pair ($i$,$j$) \\
		
		$\mathcal{N}^{\Phi}$ & Meta-path based neighbors \\
		$\mathbf{Z}_{\Phi}$ & Semantic-specific node embedding \\
		$\mathbf{q}$ & Semantic-level attention vector \\
		
		$w_{\Phi}$ & Importance of meta-path $\Phi$ \\
		$\beta_{\Phi}$ & Weight of meta-path $\Phi$ \\
		$\mathbf{Z}$ & The final embedding \\
		\bottomrule
	\end{tabular}
\end{table}

\section{The Proposed Model}
\begin{figure}
	\centering
	\includegraphics[width=1\columnwidth,height=5cm]{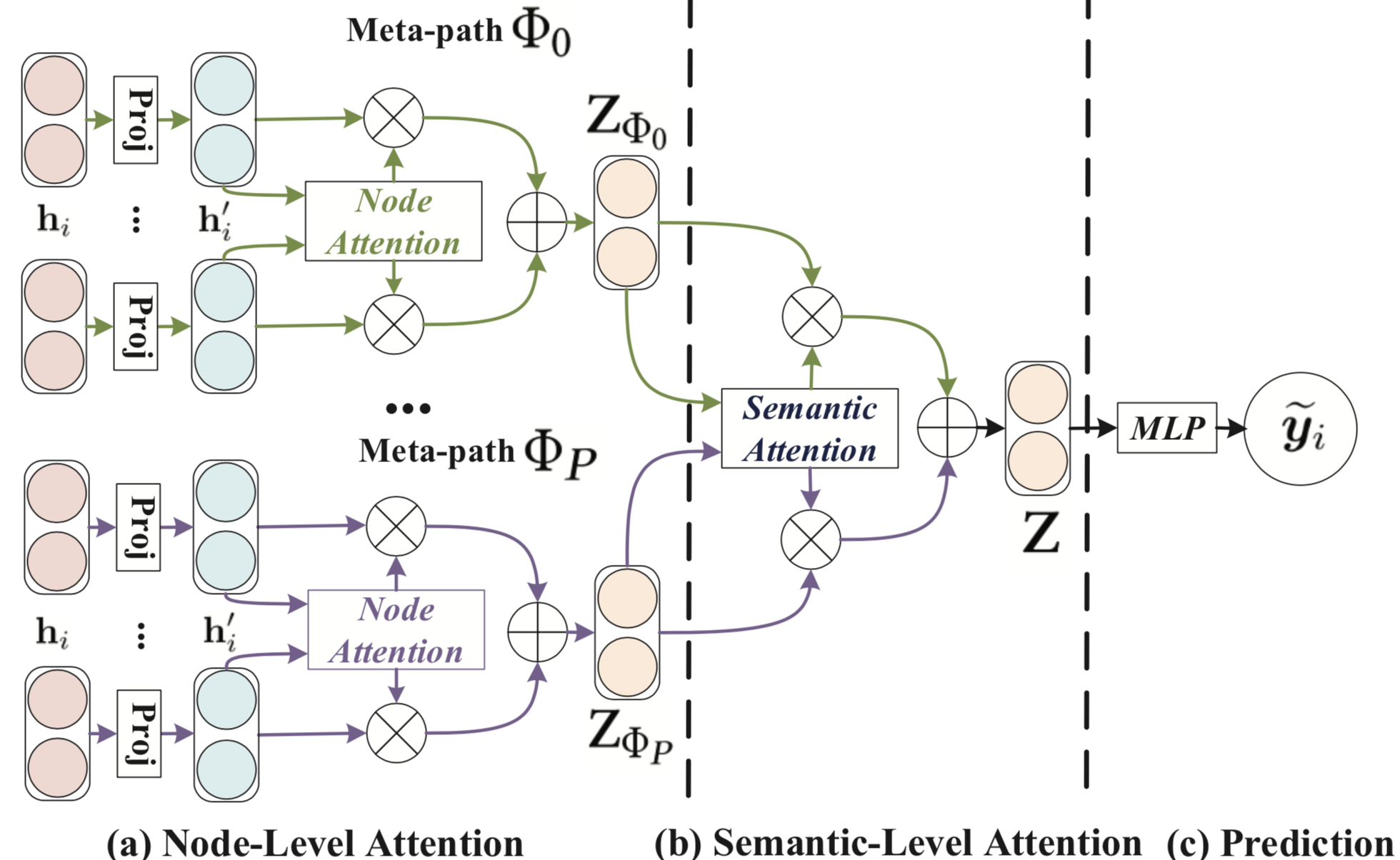}
	\caption{The overall framework of the proposed HAN. (a) All types of nodes are projected into a unified feature space and the weight of meta-path based node pair can be learned via node-level attention.  (b) Joint learning the weight of each meta-path and fuse the semantic-specific node embedding via
		semantic-level attention. 
		(c) Calculate the loss and end-to-end optimization for the proposed HAN.
	}
	\label{liuchengtu}
\end{figure}

In this section, we propose a novel semi-supervised graph neural network for heterogeneous graph.
Our model follows a hierarchical attention structure: node-level attention $\rightarrow$ semantic-level attention. Figure \ref{liuchengtu} presents the whole framework of HAN. 
First, we propose a node-level attention to learn the weight of meta-path based neighbors and aggregate them to get the semantic-specific node embedding.
After that, HAN can tell the difference of meta-paths via semantic-level attention and 
get the optimal weighted combination of  the semantic-specific node embedding for the specific task. 


\subsection{Node-level Attention}
Before aggregating the information from meta-path neighbors for each node, we should notice that the meta-path based neighbors of each node play a different role and show different importance in learning node embedding for the specific task.
Here we introduce node-level attention can learn the importance of meta-path based neighbors for each node in a heterogeneous graph and aggregate the representation of these meaningful neighbors to form a node embedding.


Due to the heterogeneity of nodes, different types of nodes have different feature spaces.
Therefore, for each type of nodes (e.g.,node with type $\phi_i$), we design the type-specific transformation matrix  $\mathbf{M}_{\phi_i}$ to project 
the features of different types of nodes into the same feature space. Unlike \cite{hamilton2018embedding}, the type-specific transformation matrix  is based on node-type rather than edge-type. The projection process can be shown as follows:
\begin{equation}
\mathbf{h}_i'= \mathbf{M}_{\phi_i} \cdot \mathbf{h}_i,\\
\end{equation}
where $\mathbf{h}_i$  and $\mathbf{h}_i'$ are the original and projected feature of node $i$, respectively. By type-specific projection operation, the node-level attention can handle arbitrary types of nodes.

After that, we leverage self-attention  \cite{attisallyouneed} to learn the weight among various 
kinds of nodes. 
Given a node pair $(i,j)$ which are connected via meta-path $\Phi$,
the node-level attention $e_{ij}^{\Phi}$ can learn the importance $e_{ij}^{\Phi}$ which means how important node $j$ will be for node $i$. The importance of meta-path based node pair $(i,j)$ can be formulated as follows:
\begin{equation}
e_{ij}^{\Phi}=att_{node}( \mathbf{h}_i', \mathbf{h}_j';\Phi).
\label{eq_2}
\end{equation}
Here $att_{node}$ denotes the deep neural network which performs the node-level attention. Given meta-path $\Phi$, $att_{node}$ is shared for all meta-path based node pairs. It is because there are some similar connection patterns under one meta-path. The above Eq. (\ref{eq_2}) shows that given meta-path $\Phi$, the weight of meta-path based node pair $(i,j)$ depends on their features.
Please note that, $e_{ij}^{\Phi}$ is asymmetric, i.e., the importance of node $i$ to node $j$ and the importance of node $j$ to node $i$ can be quite difference. It shows node-level attention can preserve the asymmetry which is a critical property of heterogenous graph.

Then we inject the structural information into the model via
masked attention
which means we only calculate the $e_{ij}^{\Phi}$ for nodes $j\in \mathcal{N}^{\Phi}_i$, where $\mathcal{N}^{\Phi}_i$ denotes the meta-path based neighbors of node $i$ (include itself). After obtaining the importance between meta-path based node pairs, we normalize them to get the weight coefficient $\alpha_{ij}^{\Phi}$ 
via softmax function:
\begin{equation}
\alpha_{ij}^{\Phi}
=softmax_j(e_{ij}^{\Phi})
=\frac{\exp \bigl(\sigma(\mathbf{a}^\mathrm{T}_{\Phi} \cdot [\mathbf{h}_i'\Vert \mathbf{h}_j'])\bigl)}{\sum_{k\in \mathcal{N}_i^{\Phi}} \exp \bigl(\sigma(\mathbf{a}^\mathrm{T}_{\Phi} \cdot [\mathbf{h}_i'\Vert \mathbf{h}_k'])\bigr)},    \\\
\label{eq3}
\end{equation}
where $\sigma$ denotes the activation function, $\Vert$ denotes the concatenate operation and 
$\mathbf{a}_{\Phi}$ is the node-level attention vector for meta-path $\Phi$. As we can see from Eq. (\ref{eq3}), the weight coefficient of $(i,j)$ depends on their features. Also please note that the weight coefficient $\alpha_{ij}^{\Phi}$ is asymmetric which means they make different contribution to each other. Not only because the concatenate order in the numerator, but also because they have different neighbors so the normalize term (denominator) will be quite difference.

\begin{figure}
	\centering
	\includegraphics[width=1\columnwidth]{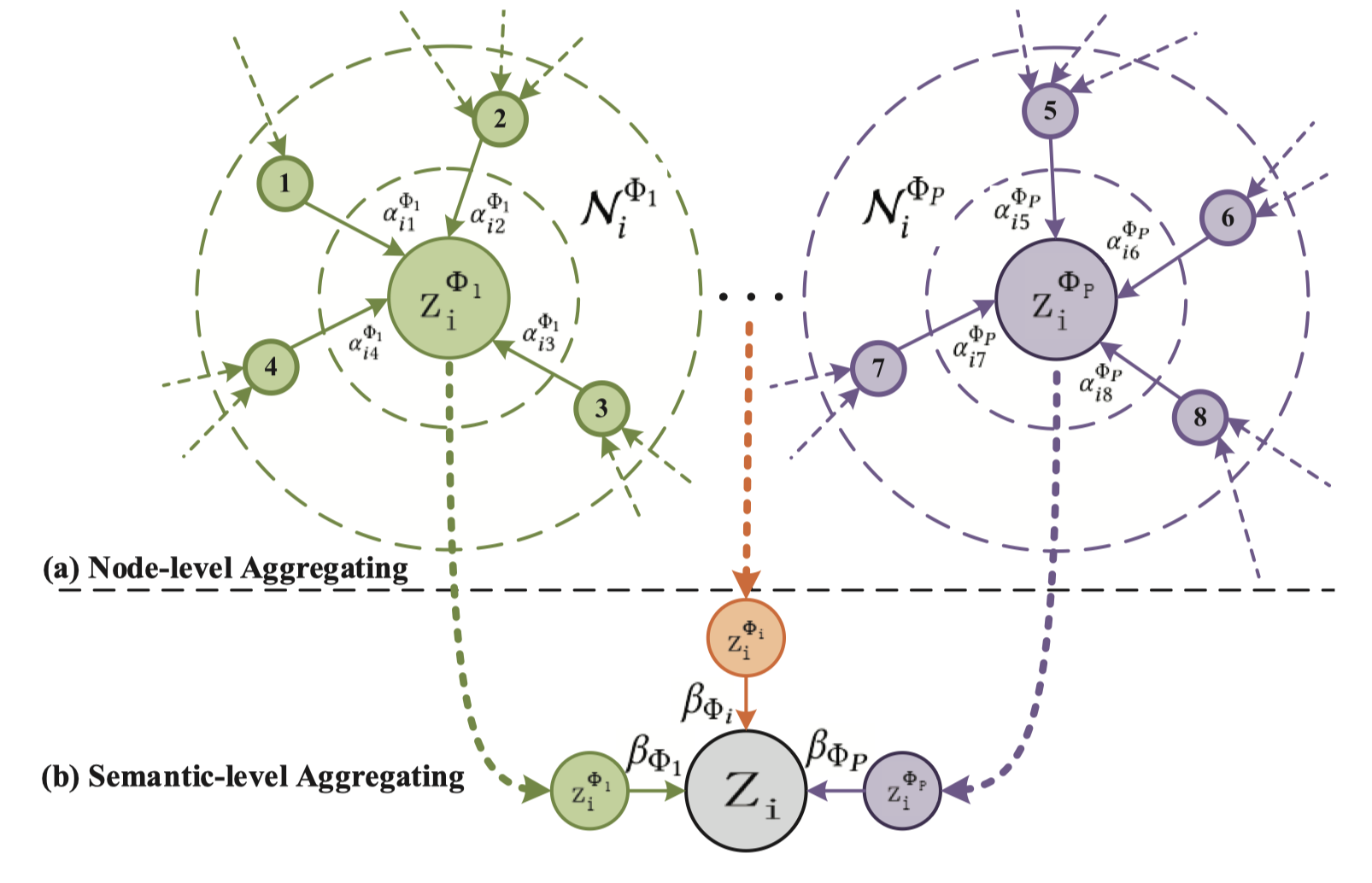}
	
	\caption{Explanation of aggregating process in both node-level and semantic-level.}
	\label{fig_explain}
\end{figure}

Then, the meta-path based embedding of node $i$ can be aggregated by the neighbor's projected features with the corresponding coefficients as follows:

\begin{equation}
\mathbf{z}^{\Phi}_i=\sigma \biggl( \sum_{j \in \mathcal{N}_i^{\Phi}} \alpha_{ij}^{\Phi} \cdot \mathbf{h}_j'  \biggr).
\label{node_agg}
\end{equation}
where $\mathbf{z}^{\Phi}_i$ is the learned embedding of node $i$ for the meta-path $\Phi$. 
To better understand the aggregating process of node-level, we also give a brief explanation in Figure \ref{fig_explain} (a). Every node embedding is aggregated by its neighors.
Since the attention weight $\alpha_{ij}^{\Phi}$ is generated for single meta-path, it is semantic-specific and able to caputre one kind of semantic information.

Since heterogeneous graph present the property of scale free,
the variance of graph data is quite high.
To tackle the above challenge,
we extend node-level attention to multihead attention so that the training process is more stable. Specifically, we repeat the node-level attention for $K$ times 
and concatenate the learned embeddings as the semantic-specific embedding:
\begin{equation}
\mathbf{z}^{\Phi}_i=
\overset{K}{\underset{k=1}{\Vert}}
\sigma \biggl( \sum_{j \in \mathcal{N}_i^{\Phi}} \alpha_{ij}^{\Phi} \cdot \mathbf{h}_j'  \biggr).
\end{equation}
Given the meta-path set  $\left\lbrace \Phi_1,\ldots,\Phi_{P}\right\rbrace $, after feeding node features into node-level attention, 
we can obtain $P$ groups of semantic-specific node embeddings, 
denoted as $\left\lbrace \mathbf{Z}_{\Phi_1},\ldots,\mathbf{Z}_{\Phi_{P}}\right\rbrace $.

\subsection{Semantic-level Attention}
Generally, every node in a heterogeneous graph contains multiple types of semantic information and semantic-specific node embedding can only reflect node
from one aspect.
To learn a more comprehensive node embedding, we need to
fuse multiple semantics which can be revealed by meta-paths. To address the challenge of meta-path selection and semantic fusion in a heterogeneous graph, we propose a novel semantic-level attention to 
automatically learn the importance of different meta-paths and fuse them for the specific task. 
Taking  $P$ groups of semantic-specific node embeddings learned from node-level attention as input, the learned weights of each meta-path $(\mathbf{\beta}_{\Phi_1},\ldots,\mathbf{\beta}_{\Phi_{P}})$ can be shown as follows:
\begin{equation}
(\mathbf{\beta}_{\Phi_1},\ldots,\mathbf{\beta}_{\Phi_{P}})=att_{sem}(\mathbf{Z}_{\Phi_1},\ldots,\mathbf{Z}_{\Phi_{P}}).
\end{equation}
Here $att_{sem}$ denotes the deep neural network which performs the semantic-level attention. 
It shows that the semantic-level attention can capture various types of semantic information behind a heterogeneous graph.

To learn the importance of each meta-path, we first transform semantic-specific embedding through a nonlinear transformation (e.g., one-layer MLP). Then we measure the importance of the semantic-specific embedding as the similarity of transformed embedding with a semantic-level attention vector $\mathbf{q}$. Furthermore, we average the importance of all the semantic-specific node embedding which can be explained as  the importance of each meta-path.
The importance of each meta-path, denoted as $w_{\Phi_i}$, is shown as follows:

\begin{equation}
w_{\Phi_p} =\frac{1}{|\mathcal{V}|}\sum_{i \in \mathcal{V}} \mathbf{q}^\mathrm{T} \cdot \tanh(\mathbf{W}\cdot \mathbf{z}_{i}^{\Phi_p}+\mathbf{b}),
\end{equation}
where $\mathbf{W}$ is the weight matrix,  $\mathbf{b}$ is the bias vector, $\mathbf{q}$ is the semantic-level attention vector. Note that for the meaningful comparation, all above parameters are shared for all meta-paths and semantic-specific embedding.
After obtaining the importance of each meta-path, we normalize them via softmax function. 
The weight of meta-path $\Phi_i$, denoted as  $\beta_{\Phi_i}$,  can be obtained by normalizing the above importance of all meta-paths using softmax function,
\begin{equation}
\beta_{\Phi_p}=\frac{\exp(w_{\Phi_p})}{\sum_{p=1}^{P} \exp(w_{\Phi_p})} ,
\end{equation}
which can be interpreted as the contribution of the meta-path $\Phi_p$ for specific task. Obviously, the higher $\beta_{\Phi_p}$, the more important meta-path $\Phi_p$ is. Note that for different tasks, meta-path $\Phi_p$ may has different weights.
With the learned weights as coefficients, 
we can fuse these semantic-specific embeddings to obtain the final embedding $\mathbf{Z}$ as follows:	
\begin{equation}
\mathbf{Z}=\sum_{p=1}^{P} \beta_{\Phi_p}\cdot \mathbf{Z}_{\Phi_p}.
\label{sem_agg}
\end{equation}
To better understand the aggregating process of semantic-level, we also give a brief explanation in Figure \ref{fig_explain} (b). The final embedding is aggregated by all semantic-specific embedding.
Then we can apply the final embedding to specific tasks and design different loss fuction. 
For semi-supervised node classification, we can minimize the Cross-Entropy over all labeled node between the ground-truth and the prediction:
\begin{equation}
L=-\sum_{l \in \mathcal{Y}_{L}} \mathbf{Y}^{l} \ln (\mathbf{C}\cdot \mathbf{Z}^{l}),
\end{equation}
where $\mathbf{C}$ is the parameter of the classifier,
$\mathcal{Y}_L$ is the set of node indices that have labels,
$\mathbf{Y}^{l}$ and $\mathbf{Z}^{l}$ are the labels and embeddings of labeled nodes.
With the guide of labeled data, we can optimize the proposed model via back propagation and learn the embeddings of nodes. The overall process of HAN in shown in Algorithm \ref{alg:aggre}.
\begin{table*}[]
	\centering
	\caption{Statistics of the datasets.}
	\label{table_datasets}
	\begin{tabular}{|c|c|c|c|c|c|c|c|c|c|}
		\hline
		Dataset               & Relations(A-B) & Number of A & Number of B & Number of A-B & Feature               & Training             & Validation           & Test                  & Meta-paths \\ \hline
		\multirow{3}{*}{DBLP} & Paper-Author   & 14328       & 4057        & 19645         & \multirow{3}{*}{334}  & \multirow{3}{*}{800} & \multirow{3}{*}{400} & \multirow{3}{*}{2857} & \emph{APA}        \\ \cline{2-5} \cline{10-10} 
		& Paper-Conf     & 14328       & 20          & 14328         &                       &                      &                      &                       & \emph{APCPA}      \\ \cline{2-5} \cline{10-10} 
		& Paper-Term     & 14327       & 8789        & 88420         &                       &                      &                      &                       & \emph{APTPA}      \\ \hline
		\multirow{2}{*}{IMDB} & Movie-Actor    & 4780        & 5841        & 14340         & \multirow{2}{*}{1232} & \multirow{2}{*}{300} & \multirow{2}{*}{300} & \multirow{2}{*}{2687} & \emph{MAM}        \\ \cline{2-5} \cline{10-10} 
		& Movie-Director & 4780        & 2269        & 4780          &                       &                      &                      &                       & \emph{MDM}        \\ \hline
		\multirow{2}{*}{ACM}  & Paper-Author   & 3025        & 5835        & 9744          & \multirow{2}{*}{1830} & \multirow{2}{*}{600} & \multirow{2}{*}{300} & \multirow{2}{*}{2125} & \emph{PAP}        \\ \cline{2-5} \cline{10-10} 
		& Paper-Subject  & 3025        & 56          & 3025          &                       &                      &                      &                       & \emph{PSP}        \\ \hline
	\end{tabular}
	
\end{table*}

\begin{algorithm}[ht]
	\SetKwInOut{Input}{\textbf{Input}}\SetKwInOut{Output}{\textbf{Output}} 
	
	\Input{
		The heterogeneous  graph $\mathcal{G} = (\mathcal{V}, \mathcal{E})$, \\
		The node feature \{$\mathbf{h}_i, \forall i\in \mathcal{V}$\},\\
		The meta-path set $ \left\lbrace \Phi_0,\Phi_1,\ldots,\Phi_{P}\right\rbrace $. \\
		The number of attention head $K$, \\
	}
	\Output{
		The final embedding $\mathbf{Z}$ ,\\
		The node-level attention weight $\alpha$ ,\\
		The semantic-level attention weight $\beta$ .\\
	}
	\BlankLine
	\For {$ \Phi_i  \in \left\lbrace \Phi_0,\Phi_1,\ldots,\Phi_{P}\right\rbrace $}{
		

		\For {$k=1...K$}{
			Type-specific transformation $\mathbf{h}_i' \leftarrow \mathbf{M}_{\phi_i} \cdot \mathbf{h}_i$ \;
			\For {$i \in \mathcal{V}$ }{
				Find the  meta-path based neighbors $\mathcal{N}^{\Phi}_i $ \;
				
				\For {$j \in \mathcal{N}^{\Phi}_i $}{
					Calculate the weight coefficient $\alpha^{\Phi}_{ij}$ \;

				}
				Calculate the semantic-specific node embedding $\mathbf{z}^{\Phi}_i \leftarrow \sigma \biggl( \sum_{j \in \mathcal{N}_i^{\Phi}} \alpha_{ij}^{\Phi} \cdot \mathbf{h}_j'  \biggr)$\;
			}
			Concatenate the learned embeddings from all attention head $\mathbf{z}^{\Phi}_i \leftarrow
			\overset{K}{\underset{k=1}{\Vert}}
			\sigma \biggl( \sum_{j \in \mathcal{N}_i^{\Phi}} \alpha_{ij}^{\Phi} \cdot \mathbf{h}_j'  \biggr)$ \;
		}
		Calculate the weight of meta-path $\beta_{\Phi_i}$ \;
		Fuse the semantic-specific embedding $\textbf{Z} \leftarrow \sum_{i=1}^{P} \beta_{\Phi_i}\cdot \textbf{Z}_{\Phi_i}$ \;
	}
	Calculate Cross-Entropy $L=-\sum_{l \in \mathcal{Y}_{L}} \mathbf{Y}_{l} \ln (\mathbf{C}\cdot \mathbf{Z}_{l})$ \;
	Back propagation and update parameters in HAN\;
	\Return{$Z, \alpha, \beta$. }
	\caption{The overall process of HAN.}
	\label{alg:aggre}
\end{algorithm}
%

\subsection{Analysis of the Proposed Model}
Here we give the analysis of the proposed HAN as follows:

	\textbullet\ The proposed model can deal with various types of nodes and relations and fuse rich semantics in a heterogeneous graph. The information can transfer from one kind of nodes to another kind of nodes via diverse relation.  Benefitted from such a heterogeneous graph attention network, different types of node embedding can enhance the mutual integration, mutual promotion and mutual upgrade. 
	
\textbullet\
	The proposed HAN is highly efficient and can be easily parallelized.
	%
	The computation of attention can compute individually across all nodes and meta-paths.
	Given a meta-path $\Phi$, the time complexity of 
	node-level attention is $O(V_{\Phi} F_1F_2 K+ E_{\Phi}F_1K)$, 
	where $K$ is the number of attention head,
	$V_{\Phi}$ is the number of nodes,  
	$E_{\Phi}$ is the number of meta-path based node pairs, 
	$F_1$ and $F_2$ are the numbers of rows and columns of the transformation matrix, respectively.
	The overall complexity is linear to the number of nodes and 
	meta-path based node pairs. The proposed model can be easily parallelized, because the node-level and semantic-level attention can be parallelized across node paris and meta-paths, respectively.
	The overall complexity is linear to the number of nodes and meta-path based node pairs.
	
\textbullet\
	The hierarchical attention is shared for the whole heterogeneous graph 
	which means the number of parameters is not 
	dependent on the scale of a heterogeneous graph and can be used for the inductive problems \cite{graphsage}. 
	Here inductive means the model  can generate node embeddings for previous unseen nodes or even unseen graph. 
	
\textbullet\
	The proposed model has potentionally good interpretability for the learned node embedding which is a big advantage for heterogeneous graph analysis.
	With the learned importance of nodes and meta-paths, 
	the proposed model can pay more attention to some meaningful nodes or meta-paths for the specific task and give a more comprensive description of a heterogeneous graph. Based on the attention values, we can check which nodes or meta-paths make the higher (or lower) contributions for our task, which is beneficial to analyze and explain our results.

\begin{table*}[]
	\centering
	\caption{Qantitative results (\%) on the node classification task.}\label{table_fenlei}
	\begin{tabular}
		{|c|c|c||c|c|c|c|c|c||c|c|c|}
		
		\hline
		Datasets                  & Metrics                   & Training & {DeepWalk}    & ESim  & {metapath2vec} & HERec & GCN   & {GAT}   & HAN$_{nd}$ & {HAN$_{sem}$}    & HAN           \\ \hline
		\multirow{8}{*}{ACM}  & \multirow{4}{*}{Macro-F1} & 20\%  & 77.25 & 77.32 & 65.09  & 66.17 & 86.81 & 86.23 & 88.15       & 89.04          & \textbf{89.40} \\
		&                           & 40\%  & 80.47 & 80.12 & 69.93  & 70.89 & 87.68 & 87.04 & 88.41       & 89.41          & \textbf{89.79} \\
		&                           & 60\%  & 82.55 & 82.44 & 71.47  & 72.38 & 88.10 & 87.56 & 87.91       & \textbf{90.00} & 89.51          \\
		&                           & 80\%  & 84.17 & 83.00 & 73.81  & 73.92 & 88.29 & 87.33 & 88.48       & 90.17          & \textbf{90.63} \\ \cline{2-12} 
		& \multirow{4}{*}{Micro-F1} & 20\%  & 76.92 & 76.89 & 65.00  & 66.03 & 86.77 & 86.01 & 87.99       & 88.85          & \textbf{89.22} \\
		&                           & 40\%  & 79.99 & 79.70 & 69.75  & 70.73 & 87.64 & 86.79 & 88.31       & 89.27          & \textbf{89.64} \\
		&                           & 60\%  & 82.11 & 82.02 & 71.29  & 72.24 & 88.12 & 87.40 & 87.68       & \textbf{89.85} & 89.33          \\
		&                           & 80\%  & 83.88 & 82.89 & 73.69  & 73.84 & 88.35 & 87.11 & 88.26       & 89.95          & \textbf{90.54} \\ \hline
		\multirow{8}{*}{DBLP} & \multirow{4}{*}{Macro-F1} & 20\%  & 77.43 & 91.64 & 90.16  & 91.68 & 90.79 & 90.97 & 91.17       & 92.03          & \textbf{92.24} \\
		&                           & 40\%  & 81.02 & 92.04 & 90.82  & 92.16 & 91.48 & 91.20 & 91.46       & 92.08          & \textbf{92.40} \\
		&                           & 60\%  & 83.67 & 92.44 & 91.32  & 92.80 & 91.89 & 90.80 & 91.78       & 92.38          & \textbf{92.80} \\
		&                           & 80\%  & 84.81 & 92.53 & 91.89  & 92.34 & 92.38 & 91.73 & 91.80       & 92.53          & \textbf{93.08} \\ \cline{2-12} 
		& \multirow{4}{*}{Micro-F1} & 20\%  & 79.37 & 92.73 & 91.53  & 92.69 & 91.71 & 91.96 & 92.05       & 92.99          & \textbf{93.11} \\
		&                           & 40\%  & 82.73 & 93.07 & 92.03  & 93.18 & 92.31 & 92.16 & 92.38       & 93.00          & \textbf{93.30} \\
		&                           & 60\%  & 85.27 & 93.39 & 92.48  & 93.70 & 92.62 & 91.84 & 92.69       & 93.31          & \textbf{93.70} \\
		&                           & 80\%  & 86.26 & 93.44 & 92.80  & 93.27 & 93.09 & 92.55 & 92.69       & 93.29          & \textbf{93.99} \\ \hline
		\multirow{8}{*}{IMDB} & \multirow{4}{*}{Macro-F1} & 20\%  & 40.72 & 32.10 & 41.16  & 41.65 & 45.73 & 49.44 & 49.78       & \textbf{50.87} & 50.00          \\
		&                           & 40\%  & 45.19 & 31.94 & 44.22  & 43.86 & 48.01 & 50.64 & 52.11       & 50.85          & \textbf{52.71} \\
		&                           & 60\%  & 48.13 & 31.68 & 45.11  & 46.27 & 49.15 & 51.90 & 51.73       & 52.09          & \textbf{54.24} \\
		&                           & 80\%  & 50.35 & 32.06 & 45.15  & 47.64 & 51.81 & 52.99 & 52.66       & 51.60          & \textbf{54.38} \\ \cline{2-12} 
		& \multirow{4}{*}{Micro-F1} & 20\%  & 46.38 & 35.28 & 45.65  & 45.81 & 49.78 & 55.28 & 54.17       & 55.01          & \textbf{55.73} \\
		&                           & 40\%  & 49.99 & 35.47 & 48.24  & 47.59 & 51.71 & 55.91 & 56.39       & 55.15          & \textbf{57.97} \\
		&                           & 60\%  & 52.21 & 35.64 & 49.09  & 49.88 & 52.29 & 56.44 & 56.09       & 56.66          & \textbf{58.32} \\
		&                           & 80\%  & 54.33 & 35.59 & 48.81  & 50.99 & 54.61 & 56.97 & 56.38       & 56.49          & \textbf{58.51} \\ \hline
	\end{tabular}
\end{table*}


\section{Experiments}
\subsection{Datasets}
The detailed descriptions of the heterogeneous graph used here are shown in Table \ref{table_datasets}.

\textbullet\	\textbf{DBLP\footnote{https://dblp.uni-trier.de}.} We
	extract a subset of DBLP which contains 14328 papers (P), 4057 authors (A), 20 conferences (C), 8789 terms (T). 
	The authors are divided into four areas: \emph{database, data mining, machine learning, information retrieval}.
	Also, we 
	label each author's research area according to the conferences they submitted. 
	Author features are the elements of a bag-of-words represented of keywords.
	Here we employ the
	meta-path set \{\emph{APA}, \emph{APCPA}, \emph{APTPA}\} to perform experiments.

\textbullet\
	\textbf{ACM\footnote{http://dl.acm.org/}.} We extract papers published in KDD, 
	SIGMOD, SIGCOMM, MobiCOMM, and VLDB and divide the papers into three classes (\emph{Database, Wireless Communication, Data Mining}). Then we construct a heterogeneous graph that comprises 3025 papers (P), 5835 authors (A) and 
	56 subjects (S). 
	Paper features correspond to elements of a bag-of-words represented of keywords.
	We employ the meta-path set \{\emph{PAP}, \emph{PSP}\} to perform experiments.
	Here we label the papers according to the conference they published.

\textbullet\
	\textbf{IMDB.} 
	Here we extract a subset of IMDB which contains 4780 movies (M), 5841 actors (A) and 2269 directors (D). 
	The movies are divided into three classes (\emph{Action, Comedy, Drama}) according to their genre. Movie features correspond to elements of a bag-of-words represented of plots.
	We employ the meta-path set \{\emph{MAM}, \emph{MDM}\} to perform experiments.

\subsection{Baselines}

We compare with some state-of-art baselines, include the (heterogeneous) network embedding methods and graph neural network based methods, to verfify the effectiveness of the proposed HAN. 
To verify the effectiveness of our node-level attention and semantic-level attention, respectively,
we also test two variants of HAN. 

\textbullet\  DeepWalk \cite{perozzi2014deepwalk}:
	A random walk based network embedding method which designs for the homogeneous graphs. 
	Here we ignore the heterogeneity of nodes and perform DeepWalk on the whole heterogeneous graph. 
	
\textbullet\   ESim \cite{Shang2016MetaPathGE}: 
	A 
	heterogeneous graph embedding method which can capture semantic information from multiple meta-paths. 
	Because it is difficult to search the weights of a set of meta-paths, we assign the weights learned from HAN to ESim. 
	
\textbullet\  metapath2vec  \cite{Dong2017metapath2vecSR}: 
	A 
	heterogeneous graph
	embedding method which performs meta-path based random walk and utilizes skip-gram to embed the heterogeneous graphs. 
	Here we test all the meta-paths for metapath2vec and report the best performance.
	
	
\textbullet\  HERec \cite{HERec}:
	A heterogeneous graph embedding method which designs a type constraint strategy to filter the node sequence
	and utilizes skip-gram to embed the heterogeneous graphs. 
	Here we test all the meta-paths for HERec and report the best performance.
	
	
\textbullet\   GCN \cite{gcn}:
	It is a semi-supervised graph convolutional network that designed for the homogeneous graphs. 
	Here we test all the meta-paths for GCN and report the best performance.
	
	
\textbullet\   GAT \cite{gat}:
	It is a semi-supervised neural network which considers the attention mechanism on the homogeneous graphs. 
	Here we test all the meta-paths for GAT and report the best performance.
	
	%

\textbullet\   HAN$_{nd}$  : 
	It is a variant of HAN, which removes node-level attention and assigns the same importance to each neighbor.
	
\textbullet\  HAN$_{sem}$  : 
	It is a variant of HAN, which removes the semantic-level attention and assigns  the same importance to each meta-path.
	
\textbullet\  HAN: 
	The proposed semi-supervised graph neural network
	which employs node-level attention and semantic-level attention simultaneously.

\subsection{Implementation Details}
For the proposed HAN, we 
randomly initialize parameters
and optimize the 
model with Adam
\cite{adam}.
For the proposed HAN, we set the learning rate to 0.005, 
the regularization parameter to 0.001, the dimension of the semantic-level attention vector $\mathbf{q}$ to 128, the number of attention head $K$ to 8, the dropout of attention to 0.6.
And we use early stopping with a patience of 100, i.e. we stop training if the validation loss does not decrease for 100 consecutive epochs. To make our experiments repeatable, we make our dataset and codes publicly available at website\footnote{https://github.com/Jhy1993/HAN}.For GCN and GAT, we optimize their parameters using the validation set. 
For semi-supervised graph neural network, including GCN, GAT and HAN, we split exactly the same training set, validation set and test set to ensure fairness.
For random walk based methods include DeepWalk, ESim, metapath2vec, and HERec, 
we set window size to 5, walk length to 100, 
walks per node to 40, the number of negative samples to 5. For a fair comparison, we set the embedding dimension to 64 for all the above algorithms. 

\subsection{Classification}
Here we employ 
KNN classifier with ${k=5}$ to perform node classification. 
Since the variance of graph-structured data can be quite high, we repeat the process for 10 times and report the averaged \emph{Macro-F1} and \emph{Micro-F1} in Table \ref{table_fenlei}. 

Based on Table \ref{table_fenlei}, we can see that HAN achieves the best performance. 
For traditional heterogeneous graph embedding method, ESim which can leverage multiple meta-paths performs better than metapath2vec.
Generally, graph neural network based methods which combine the structure and feature information, e.g., GCN and GAT, usually perform better. 
To go deep into these methods, compared to simply average over node neighbors, e.g., GCN and HAN$_{nd}$, 
GAT and HAN can weigh the information properly and improve the performance of the learned embedding. 
Compared to GAT, the proposed HAN, which designs for heterogeneous graph, captures the rich semantics successfully and shows its superiority.
Also, without node-level attention (HAN$_{nd}$) or semantic-level attention (HAN$_{sem}$), 
the performance becomes worse than HAN, which indicates the importance of modeling the attention mechanism on both of the nodes and semantics. Note that in ACM and IMDB, HAN improves classification results more significantly than in DBLP. Mainly because \emph{APCPA} is the much more important than the rest meta-paths.
We will explain this phenomenon in Section 5.7 by analyzing the semantic-level attention.

Through the above analysis, we can find that the proposed HAN achieves the best performance on all datasets. The results demonstrate that it is quite important to capture the importance of nodes and meta-paths in heterogeneous graph analysis.

\begin{table*}[]
	\centering
	\caption{Qantitative results (\%) on the node clustering task.}
	\label{table_julei}
	\begin{tabular}{|c|c||c|c|c|c|c|c||c|c|c|}
		\hline
		Datasets              & Metrics & DeepWalk    & ESim  & metapath2vec & HERec & GCN   & GAT   & HAN$_{nd}$ & HAN$_{sem}$ & HAN           \\ \hline
		\multirow{2}{*}{ACM}  & NMI     & 41.61 & 39.14 & 21.22  & 40.70 & 51.40 & 57.29 & 60.99       & 61.05       & \textbf{61.56} \\
		& ARI     & 35.10 & 34.32 & 21.00  & 37.13 & 53.01 & 60.43 & 61.48       & 59.45       & \textbf{64.39} \\ \hline
		\multirow{2}{*}{DBLP} & NMI     & 76.53 & 66.32 & 74.30  & 76.73 & 75.01 & 71.50 & 75.30       & 77.31       & \textbf{79.12} \\
		& ARI     & 81.35 & 68.31 & 78.50  & 80.98 & 80.49 & 77.26 & 81.46       & 83.46       & \textbf{84.76} \\ \hline
		\multirow{2}{*}{IMDB} & NMI     & 1.45  & 0.55  & 1.20   & 1.20  & 5.45  & 8.45  & 9.16        & 10.31       & \textbf{10.87} \\
		& ARI     & 2.15  & 0.10  & 1.70   & 1.65  & 4.40  & 7.46  & 7.98        & 9.51       & \textbf{10.01} \\ \hline
	\end{tabular}
\end{table*}

\subsection{Clustering}
We also conduct the clustering task to evaluate the embeddings learned from the above algorithms. 
Once the proposed HAN trained, we can get all the node embedding via feed forward.
Here we utilize the KMeans to perform node clustering and the number of clusters $K$ is set to the number of classes. We use the same ground-truth as in node classification.
And we adopt \emph{NMI} and \emph{ARI} to assess the quality of the clustering results. 
Since the performance of KMeans is affected by initial centroids, we repeat the process for 10 times and report the average results in Table \ref{table_julei}.

As can be seen in Table \ref{table_julei}, we can find that HAN performs consistently much better than all baselines. Also, 
graph neural network based algorithms usually achieve better performance. 
Besides, without distinguishing the importance of nodes or meta-paths, 
metapath2vec and GCN cannot perform well.
With the guide of multiple meta-paths, HAN performs significantly better than GCN and GAT. 
On the other hand, without node-level attention (HAN$_{nd}$) or semantic-level attention (HAN$_{sem}$), the performance of HAN has shown various degrees of degeneration. It demonstrates that via assigning the different importance to nodes and meta-paths, the proposed HAN can learn a more meaningful node embedding. 

Based on the above analysis, we can find that the propsed HAN can give a comprehensive description of heterogeneous graph and achieve a significant improvements.

\begin{figure}
	\centering
	\subfigure[Meta-path based neighbors of P831]{\includegraphics[width=0.49\columnwidth]{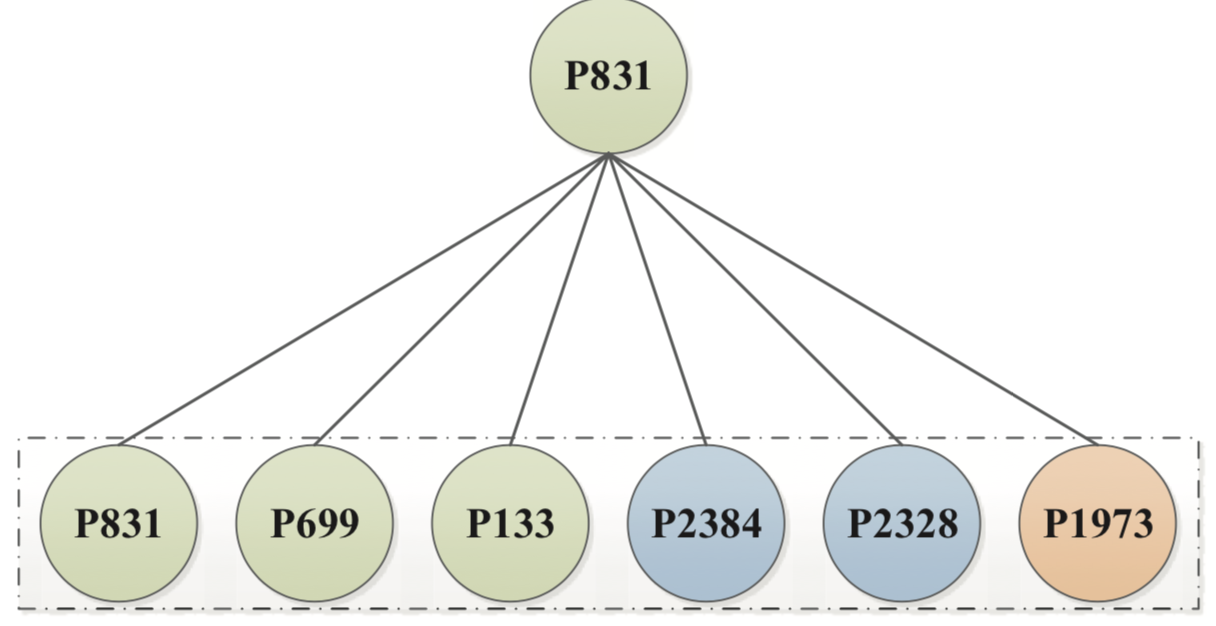}}
	\subfigure[Attention values of P831's neighbors]{\includegraphics[width=0.49\columnwidth]{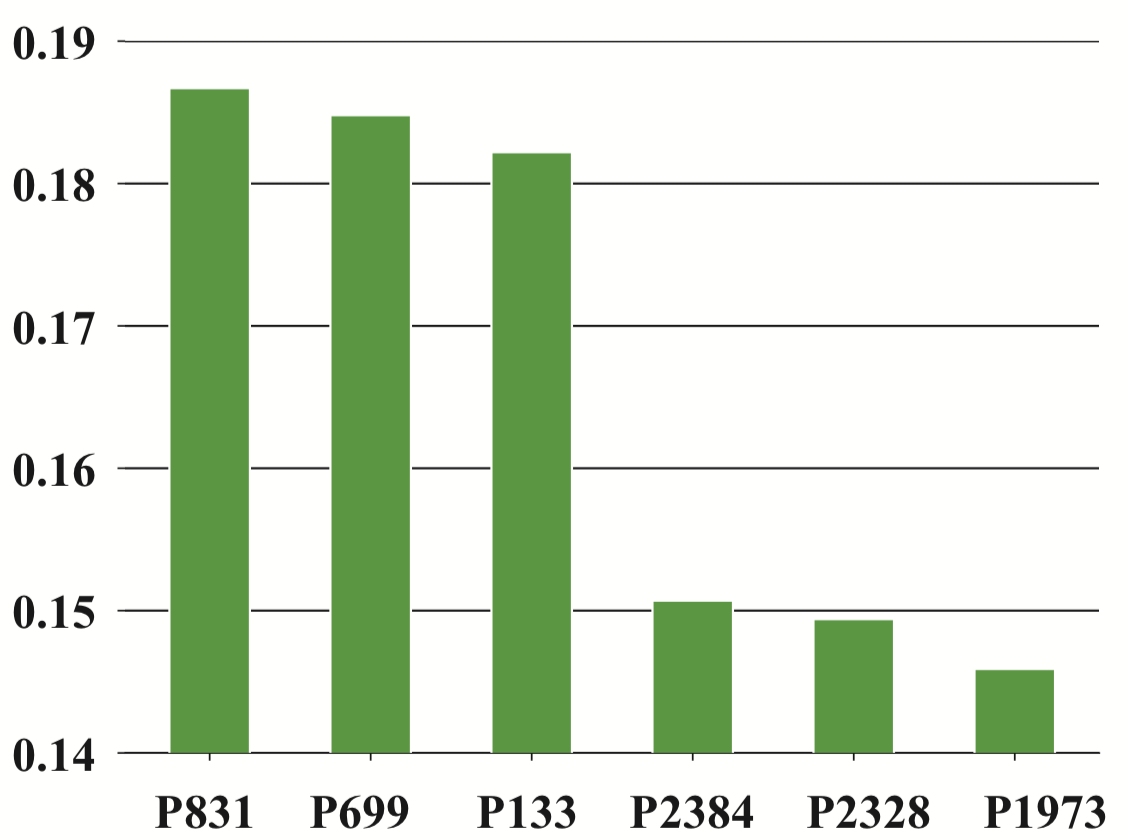}}
	\caption{Meta-path based neighbors of node P831 and corresponding attention values (Different colors mean different classes, e.g., \emph{green} means Data Mining, \emph{blue} means Database, \emph{orange} means Wireless Communication).}
	\label{fig_node_att}
\end{figure}


\begin{figure}
	\centering
	\subfigure[NMI values on DBLP]{\includegraphics[width=0.49\columnwidth]{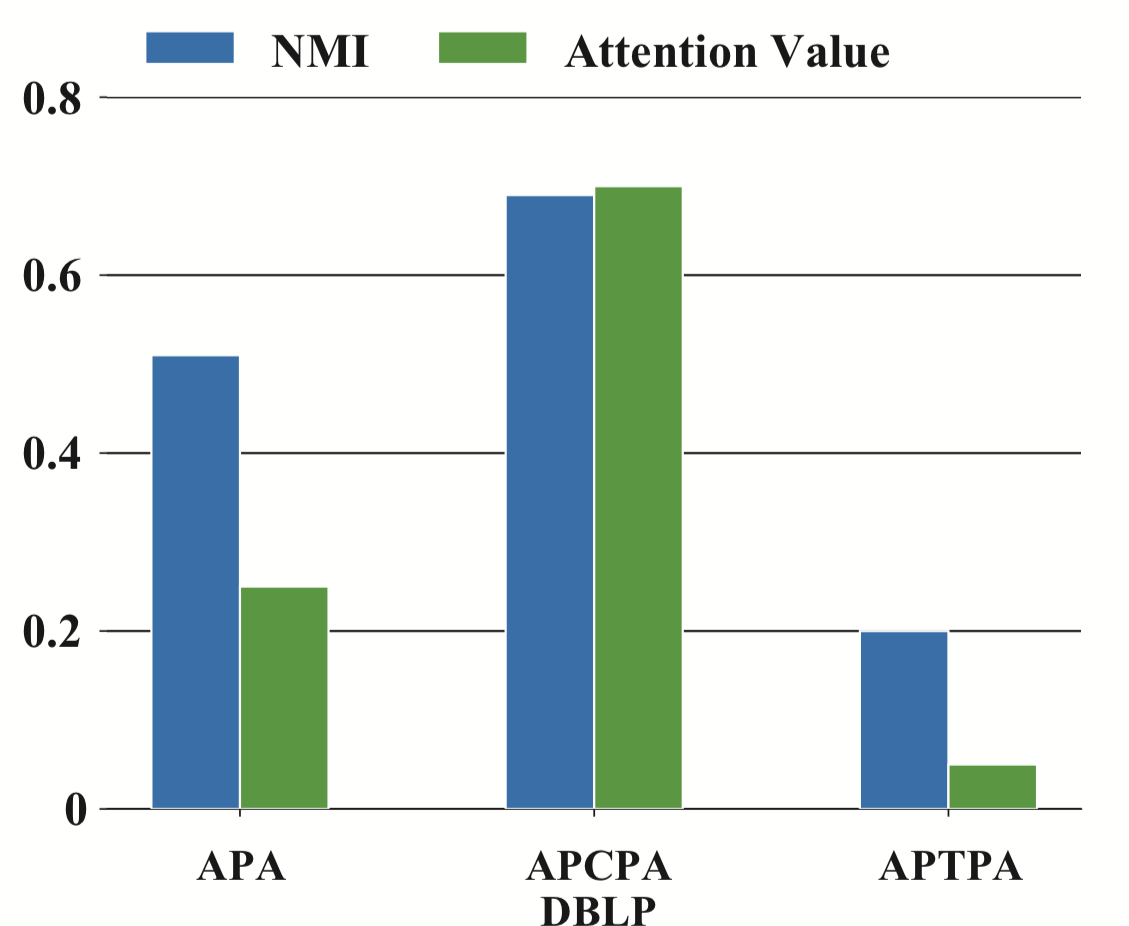}}
	\subfigure[NMI values on ACM]{\includegraphics[width=0.49\columnwidth]{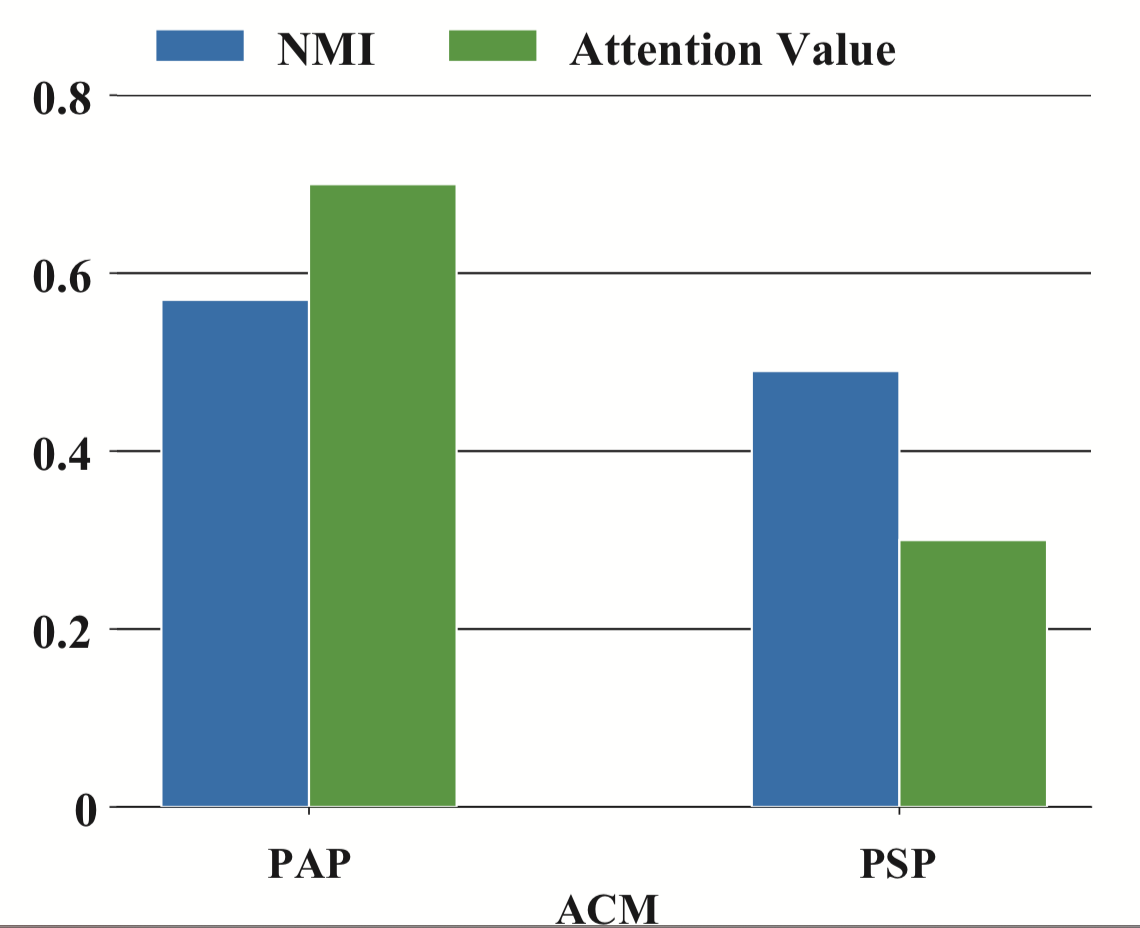}}
	\caption{Performance of single meta-path and corresponding attention value.}
	\label{fig_att_metapath}
\end{figure}

\subsection{Analysis of Hierarchical Attention Mechanism}
A salient property of HAN is the incorporation of the hierarchical mechanism, 
which takes the importance of node neighbors and meta-paths into consideration in learning representative embedding. 
Recall that we have learned the node-level attention weight $\alpha^{\Phi}_{ij}$ and the semantic-level attention weight $\beta_{\Phi_i}$. To better understand the importance of the neighbors and meta-paths, we provide a detailed analysis on the hierarchical attention mechanism.

\textbf{Analysis of node-level attention.} 
As mentioned before, given a specific task, our model can learn the attention values between nodes and its neighbors in a meta-path. Some important neighbors which are useful for the specific task tend to have larger attention values.
Here we take the paper P831 \footnote{Xintao Wu, Daniel Barbara, Yong Ye. Screening and Interpreting Multi-item Associations Based on Log-linear Modeling, KDD'03} in ACM dataset as an illustrative example. 
Given a meta-path Paper-Author-Paper which describes the co-author of different papers, we enumerate the meta-path based neighbors of paper P831 and their attention values are shown in Figure \ref{fig_node_att}. From Figure \ref{fig_node_att}(a), we can see that P831 connects to  P699 \footnote{Xintao Wu, Jianpin Fan, Kalpathi Subramanian. B-EM: a classifier incorporating bootstrap with EM approach for data mining, KDD'02} and P133 \footnote{Daniel Barbara, Carlotta Domeniconi, James P. Rogers. Detecting outliers using transduction and statistical testing, KDD'06}, which all belong to \emph{Data Mining}; conects to P2384 \footnote{Walid G. Aref, Daniel Barbara, Padmavathi Vallabhaneni. The Handwritten Trie: Indexing Electronic Ink, SIGMOD'95} and P2328 \footnote{Daniel Barbara, Tomasz Imielinski. Sleepers and Workaholics: Caching Strategies in Mobile Environments, VLDB'95} while P2384 and P2328 both belong to \emph{Database}; connects to P1973 \footnote{Hector Garcia-Holina, Daniel Barbara. The cost of data replication, SIGCOMM'81} while P1973 belongs to \emph{Wireless Communication}. From Figure \ref{fig_node_att}(b), we can see that paper P831 gets the highest attention value from node-level attention which means the node itself plays the most important role in learning its representation. It is reasonable because all information supported by neighbors are usually viewed as a kind of supplementary information. Beyond itself, P699 and P133 get the second and third largest attention values.
This is because P699 and P133 also belong to \emph{Data Mining} and they can make significant contribution to identify the class of  P831.
The rest neighbors get minor attention values that because they do not belong to \emph{Data Mining} and cannot make important contribution to identify the P831's class.
Based on the above analysis, we can see that the node-level attention can tell the difference among neighbors and assign higher weights to some meaningful neighbors. 
\begin{figure*}
	\centering
	
	\subfigure[GCN]{\includegraphics[width=0.49\columnwidth]{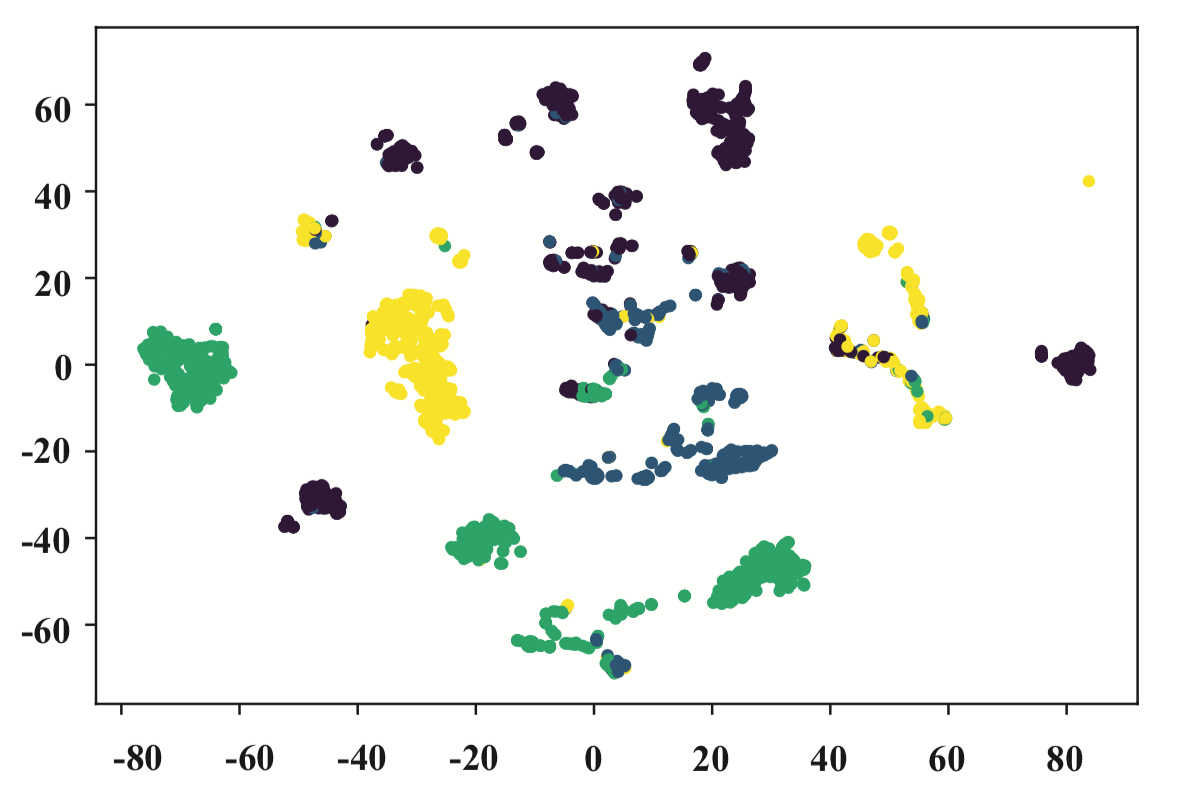}}
	\subfigure[GAT]{\includegraphics[width=0.49\columnwidth]{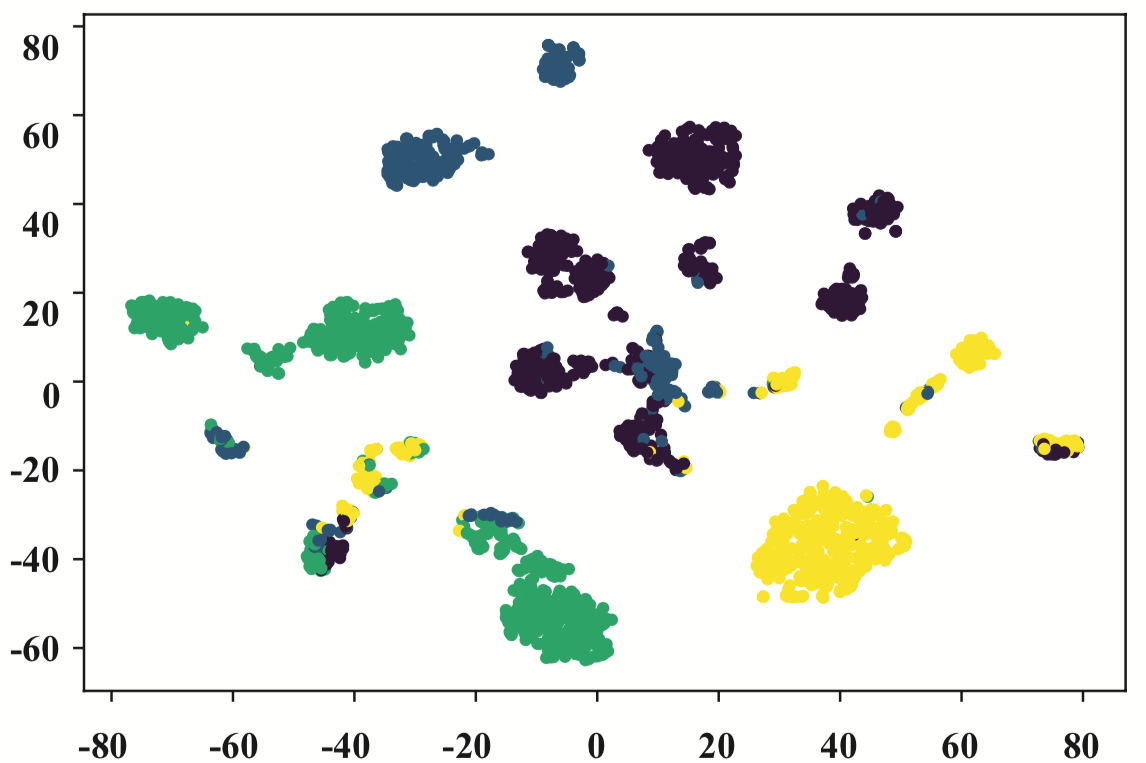}}
	\subfigure[metapath2vec]{\includegraphics[width=0.49\columnwidth]{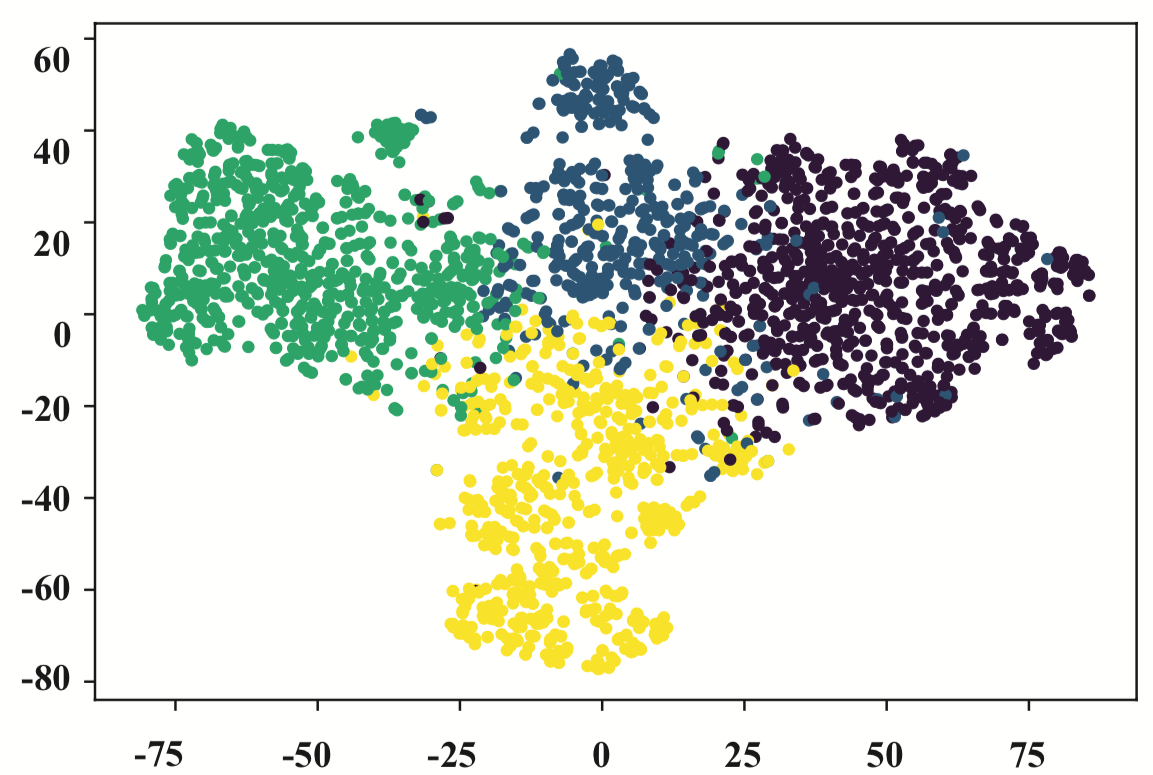}}
	\subfigure[HAN]{\includegraphics[width=0.49\columnwidth]{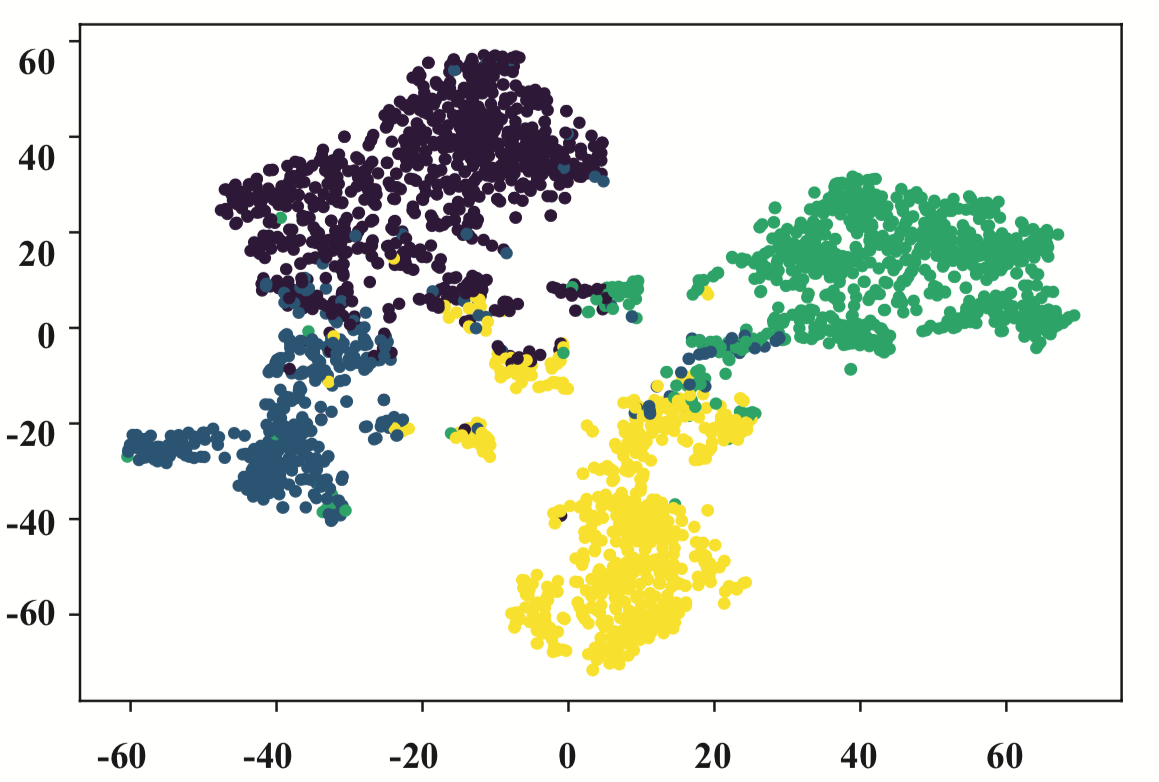}}
	\caption{Visualization embedding on DBLP. Each point indicates one author and its color indicates the research area. }
	\label{fig_vis}
\end{figure*}

%
%
%
%
%

\begin{figure*}
	\centering
	\subfigure[Dimension of the final embedding $\mathbf{Z}$]{\includegraphics[width=0.55\columnwidth,height=4cm]{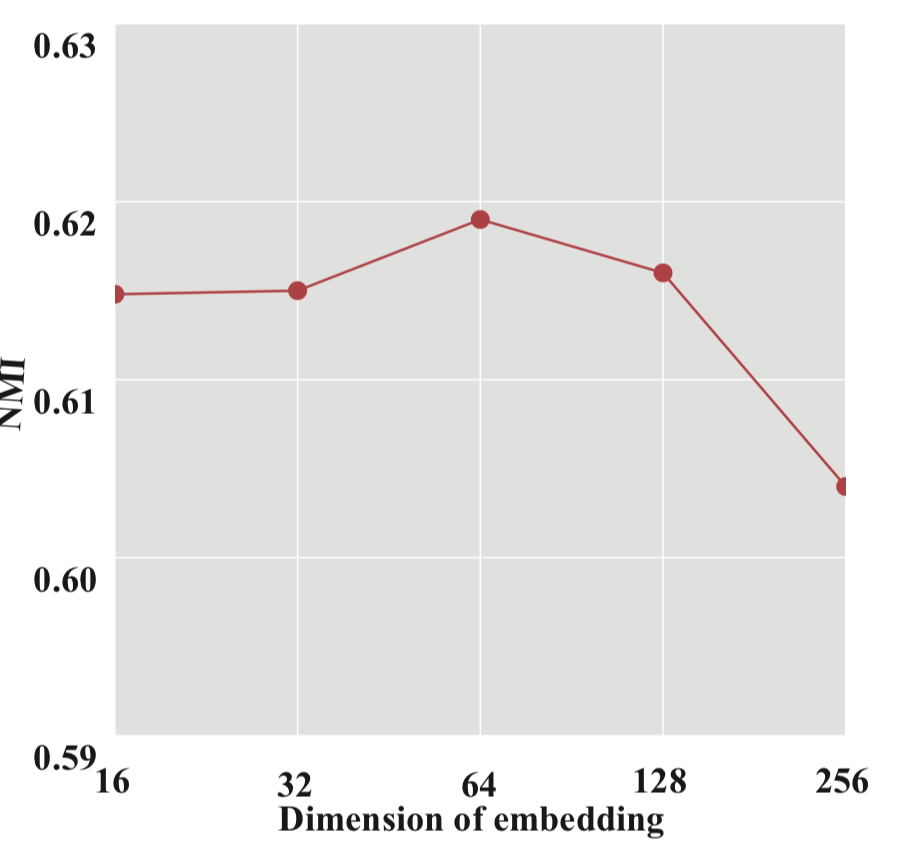}}
	\subfigure[Dimension of the semantic-level attention vector $\mathbf{q}$]{\includegraphics[width=0.55\columnwidth,height=4cm]{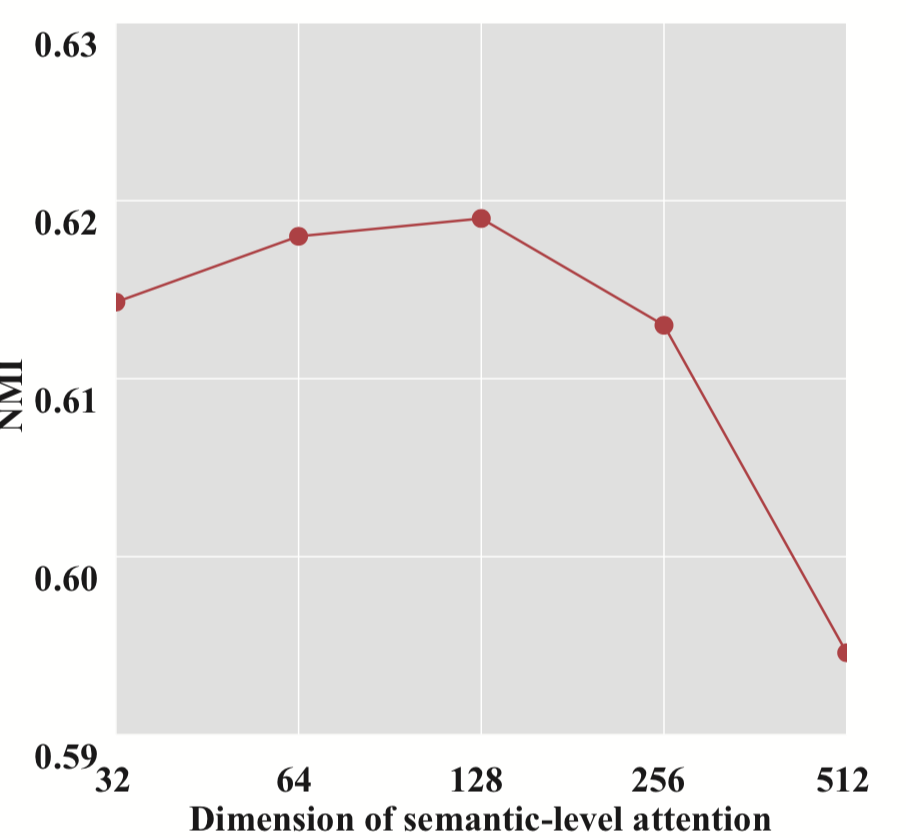}}
	\subfigure[Number of attention head $K$.]{\includegraphics[width=0.55\columnwidth,height=4cm]{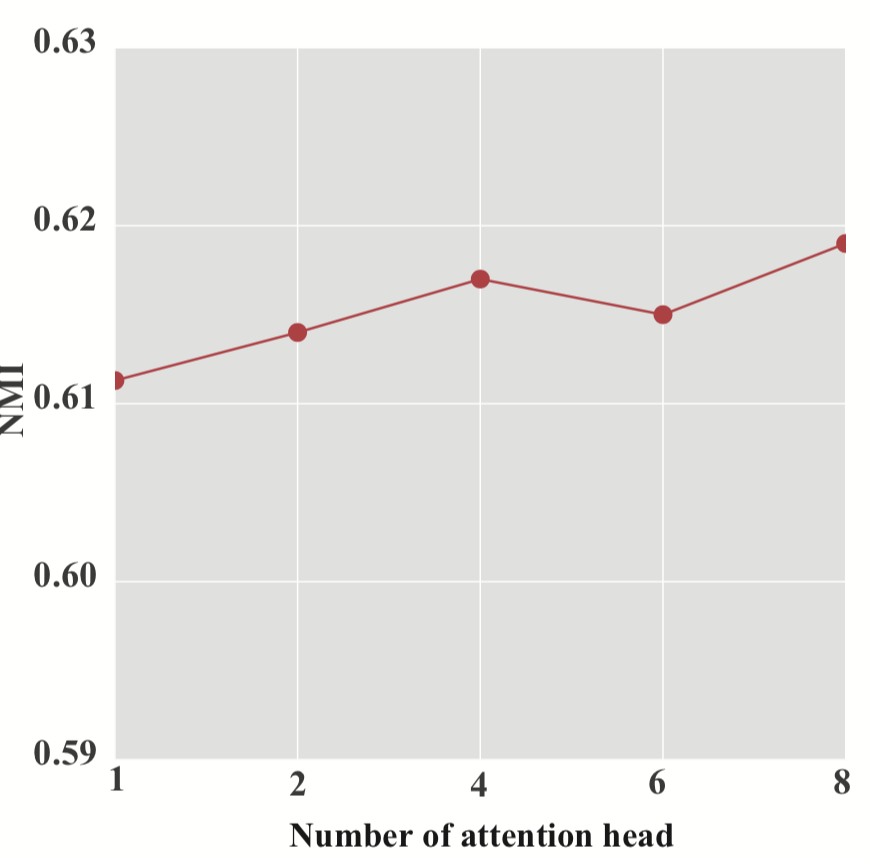}}
	\caption{Parameter sensitivity of HAN w.r.t. Dimension of the final embedding $Z$,  Dimension of the semantic-level attention vector $q$ and Number of attention head $K$.}
	\label{fig_para2}
\end{figure*}

\textbf{Analysis of semantic-level attention.} 
As mentioned before, the proposed HAN can learn the importance of meta-paths for the specific task.
To verify the ability of semantic-level attention, taking DBLP and ACM as examples, we report the clustering results (\emph{NMI}) of single meta-path and corresponding attention values in Figure \ref{fig_att_metapath}. Obviously, there is a positive correlation between the performance of a single meta-path and its attention value. 
For DBLP, HAN gives \emph{APCPA} the largest weight, which means that HAN considers the \emph{APCPA} as the most critical meta-path in identifying the author's research area. It makes sense because the author's research area and the conferences they submitted are highly correlated. For example, some natural language processing researchers mainly submit their papers to ACL or EMNLP, whereas some data mining researchers may submit their papers to KDD or WWW.
Meanwhile, it is difficult for \emph{APA} to identify the author's research area well. If we treat these meta-paths equally, e.g., HAN$_{sem}$, the performance will drop significantly. 
Based on the attention values of each meta-path, we can find that the meta-path \emph{APCPA} is much more useful than \emph{APA} and \emph{APTPA}. So even the proposed HAN can fuse them, \emph{APCPA} still plays a leading role in identifying the author's research area  while \emph{APA} and \emph{APTPA} do not.
It also explains why the performance of HAN in DBLP may not be as significant as in ACM and IMDB.
We get the similar conclusions on ACM.
For ACM, the results show that HAN gives the most considerable weight to \emph{PAP}. Since the performance of \emph{PAP} is slightly better than \emph{PSP}, so HAN$_{sem}$ can achieve good performance by simple average operation. We can see that semantic-level attention can reveal the difference between these meta-paths and weights them adequately.

\subsection{Visualization}

For a more intuitively comparation, we conduct the task of visualization,  which aims to layout a heterogeneous graph on a low dimensional space. Specifically, we learn the node embedding based on the proposed model and project the learned embedding into a 2-dimensional space.
Here we utilize t-SNE \cite{maaten2008visualizing} to visualize the author embedding in DBLP and coloured the nodes based on their research areas. 

From Figure \ref{fig_vis}, we can find that GCN and GAT which design for the homogeneous graphs do not perform well. The authors belong to different research areas are mixed with each other.
Metapath2vec performs much better than the above homogeneous graph neural networks.
It demonstrates that the proper meta-path(e.g., \emph{APCPA}) can make a significant contribution to heterogeneous graph analysis.
However, since metapath2vec can only take only one meta-path into consideration, the boundary is still blurry.
From Figure \ref{fig_vis}, we can see that the visualization of  HAN peform best. 
With the guide of multiple meta-paths,  the embedding learned by HAN has high intra-class similarity and separates the authors in different research area with distinct boundaries. 

\subsection{Parameters Experiments}
In this section, we investigate the sensitivity of parameters and report the results of clustering (\emph{NMI}) on ACM dataset with various parameters in Figure \ref{fig_para2}.

\textbullet\  \textbf{Dimension of the final embedding $\mathbf{Z}$.} We first test the effect of the dimension of the final embedding $\mathbf{Z}$. 
	The result is shown in Figure \ref{fig_para2}(a).
	We can see that with the growth of the embedding dimension, 
	the performance raises first and then starts to drop slowly. 
	The reason is that HAN needs a suitable dimension to encode the semantics information and 
	larger dimension may introduce additional redundancies. 

\textbullet\   \textbf{Dimension of semantic-level attention vector $\mathbf{q}$.} Since the ability of semantic-level attention is affected by the dimension of the semantic-level attention vector $\mathbf{q}$, we explore the experimental results 
	with various dimension. The result is shown in Figure \ref{fig_para2}(b).
	We can find that the performance of HAN grows with the dimension of semantic-level attention vector and achieves the best performance when the dimension of $\mathbf{q}$ is set to 128.
	After that, the performance of HAN starts to degenerate which may because of overfitting.
	

\textbullet\  \textbf{Number of attention head $K$.} In order to check the impact of multihead attention, we explore the performance of HAN with various number of attention head. The result is shown in Figure \ref{fig_para2}(c). Note that the multihead attention is removed when the number of attention head is set to 1. Based on the results, we can find that the more number of attention head will generally improve the performance of HAN. However, with the change of attention head, the performance of HAN improve only slightly.
	Meanwhile, we also find that multihead attention can make the training process more stable.



\section{Conclusion}
In this paper, we tackle several fundamental problems in heterogeneous graph analysis and propose a semi-supervised heterogeneous graph neural network based solely on attention mechanism.
The proposed HAN can capture complex structures and rich semantics behind heterogeneous graph.
The proposed model leverages node-level attention and semantic-level attention to learn the importance of nodes and meta-paths, respectively. 
Meanwhile, the proposed model utilizes the structural information and the feature information in a uniform way.
Experimental results include classification and clustering 
demonstrate the effectiveness of HAN. By analyzing the learned attention weights include both node-level and semantic-level, 
the proposed HAN has proven its potentially good interpretability.


\section{Acknowledgments}
This work is supported in part by the National Natural Science Foundation of China (No. 61702296, 61772082,  61532006), the Beijing Municipal Natural Science Foundation (4182043), and the CCF-Tencent Open Fund.

\bibliographystyle{ACM-Reference-Format}
\bibliography{han}

\end{document}